\pdfoutput=1

\documentclass[12pt,a4paper]{article}

\usepackage{microtype}
\usepackage{lineno}  
\usepackage{graphicx}  
\usepackage{xspace} 

\usepackage{color}
\usepackage{soul}
\usepackage{colortbl}
\usepackage{amsmath} 
\usepackage{ifthen} 
\graphicspath{{./figs/}} 

\newboolean{pdflatex}
\setboolean{pdflatex}{true} 
\newboolean{articletitles}
\setboolean{articletitles}{true} 

\newboolean{uprightparticles}
\setboolean{uprightparticles}{false} 

\usepackage{amssymb}
\usepackage{amsfonts}
\usepackage{upgreek} 

\newcommand*\patchAmsMathEnvironmentForLineno[1]{%
\expandafter\let\csname old#1\expandafter\endcsname\csname #1\endcsname
\expandafter\let\csname oldend#1\expandafter\endcsname\csname
end#1\endcsname
 \renewenvironment{#1}%
   {\linenomath\csname old#1\endcsname}%
   {\csname oldend#1\endcsname\endlinenomath}%
}
\newcommand*\patchBothAmsMathEnvironmentsForLineno[1]{%
  \patchAmsMathEnvironmentForLineno{#1}%
  \patchAmsMathEnvironmentForLineno{#1*}%
}
\AtBeginDocument{%
\patchBothAmsMathEnvironmentsForLineno{equation}%
\patchBothAmsMathEnvironmentsForLineno{align}%
\patchBothAmsMathEnvironmentsForLineno{flalign}%
\patchBothAmsMathEnvironmentsForLineno{alignat}%
\patchBothAmsMathEnvironmentsForLineno{gather}%
\patchBothAmsMathEnvironmentsForLineno{multline}%
}

\usepackage{hyperref}    
\usepackage[all]{hypcap} 

\def\lhcb {\mbox{LHCb}\xspace}
\def\ux85 {\mbox{UX85}\xspace}

\ifthenelse{\boolean{uprightparticles}}%
{

 \def\Pphi        {\ensuremath{\upphi}\xspace}

 \def\Ppsi        {\ensuremath{\uppsi}\xspace}

 \def\PDelta      {\ensuremath{\Delta}\xspace}                 
 \def\PXi      {\ensuremath{\Xi}\xspace}                 
 \def\PLambda      {\ensuremath{\Lambda}\xspace}                 
 \def\PSigma      {\ensuremath{\Sigma}\xspace}                 
 \def\POmega      {\ensuremath{\Omega}\xspace}                 
 \def\PUpsilon      {\ensuremath{\Upsilon}\xspace}                 
 
 \def\PB      {\ensuremath{\mathrm{B}}\xspace}                 
                  
 \def\PD      {\ensuremath{\mathrm{D}}\xspace}

 \def\PJ      {\ensuremath{\mathrm{J}}\xspace}                 
 \def\PK      {\ensuremath{\mathrm{K}}\xspace}

 \def\Pb      {\ensuremath{\mathrm{b}}\xspace}                 
 \def\Pc      {\ensuremath{\mathrm{c}}\xspace}

 \def\Pi      {\ensuremath{\mathrm{i}}\xspace}

 \def\Ps      {\ensuremath{\mathrm{s}}\xspace}

}
{

 \def\Pphi        {\ensuremath{\phi}\xspace}

 \def\Ppsi        {\ensuremath{\psi}\xspace}                 
                  
 \mathchardef\PDelta="7101
 \mathchardef\PXi="7104
 \mathchardef\PLambda="7103
 \mathchardef\PSigma="7106
 \mathchardef\POmega="710A
 \mathchardef\PUpsilon="7107
                  
 \def\PB      {\ensuremath{B}\xspace}                 
                  
 \def\PD      {\ensuremath{D}\xspace}

 \def\PJ      {\ensuremath{J}\xspace}                 
 \def\PK      {\ensuremath{K}\xspace}

 \def\Pb      {\ensuremath{b}\xspace}                 
 \def\Pc      {\ensuremath{c}\xspace}

 \def\Pi      {\ensuremath{i}\xspace}

 \def\Ps      {\ensuremath{s}\xspace}

}

\def\squark    {\ensuremath{\Ps}\xspace}

\def\cquark    {\ensuremath{\Pc}\xspace}

\def\bquark    {\ensuremath{\Pb}\xspace}

\def\kaon  {\ensuremath{\PK}\xspace}
  \def\Kbar  {\kern 0.2em\overline{\kern -0.2em \PK}{}\xspace}

\def\Kz    {\ensuremath{\kaon^0}\xspace}
\def\Kzb   {\ensuremath{\Kbar^0}\xspace}
\def\KzKzb {\ensuremath{\Kz \kern -0.16em \Kzb}\xspace}
\def\Kp    {\ensuremath{\kaon^+}\xspace}
\def\Km    {\ensuremath{\kaon^-}\xspace}

\def\KpKm  {\ensuremath{\Kp \kern -0.16em \Km}\xspace}

\def\Kstarz  {\ensuremath{\kaon^{*0}}\xspace}
\def\Kstarzb {\ensuremath{\Kbar^{*0}}\xspace}

  \def\Dbar    {\kern 0.2em\overline{\kern -0.2em \PD}{}\xspace}
\def\D       {\ensuremath{\PD}\xspace}
\def\Db      {\ensuremath{\Dbar}\xspace}
\def\Dz      {\ensuremath{\D^0}\xspace}
\def\Dzb     {\ensuremath{\Dbar^0}\xspace}
\def\DzDzb   {\ensuremath{\Dz {\kern -0.16em \Dzb}}\xspace}
\def\Dp      {\ensuremath{\D^+}\xspace}
\def\Dm      {\ensuremath{\D^-}\xspace}

\def\DpDm    {\ensuremath{\Dp {\kern -0.16em \Dm}}\xspace}

\def\Ds      {\ensuremath{\D^+_\squark}\xspace}
\def\Dsp     {\ensuremath{\D^+_\squark}\xspace}

\def\B       {\ensuremath{\PB}\xspace}
  \def\Bbar    {\kern 0.18em\overline{\kern -0.18em \PB}{}\xspace}
\def\Bb      {\ensuremath{\Bbar}\xspace}

\def\Bzb     {\ensuremath{\Bbar^0}\xspace}
\def\Bu      {\ensuremath{\B^+}\xspace}

\def\Bp      {\ensuremath{\Bu}\xspace}

\def\Bsb     {\ensuremath{\Bbar^0_\squark}\xspace}

\def\jpsi     {\ensuremath{{\PJ\mskip -3mu/\mskip -2mu\Ppsi\mskip 2mu}}\xspace}

  \def\Y#1S{\ensuremath{\PUpsilon{(#1S)}}\xspace}

\def\Lbar {\ensuremath{\kern 0.1em\overline{\kern -0.1em\Lambda\kern -0.05em}\kern 0.05em{}}\xspace}

\def\BF         {{\ensuremath{\cal B}\xspace}}

\def\BR         {\BF}
\newcommand{\decay}[2]{\ensuremath{#1\!\to #2}\xspace}         

\def\to                 {\ensuremath{\rightarrow}\xspace}

\def\CP                {\ensuremath{C\!P}\xspace}

\def\AT#1     {\ensuremath{A_{\mathrm{T}}^{#1}}\xspace}           

\def\C#1      {\ensuremath{\mathcal{C}_{#1}}\xspace}                       
\def\Cp#1     {\ensuremath{\mathcal{C}_{#1}^{'}}\xspace}                    
\def\Ceff#1   {\ensuremath{\mathcal{C}_{#1}^{\mathrm{(eff)}}}\xspace}        
\def\Cpeff#1  {\ensuremath{\mathcal{C}_{#1}^{'\mathrm{(eff)}}}\xspace}       
\def\Ope#1    {\ensuremath{\mathcal{O}_{#1}}\xspace}                       
\def\Opep#1   {\ensuremath{\mathcal{O}_{#1}^{'}}\xspace}                    

\newcommand{\tev}{\ensuremath{\mathrm{\,Te\kern -0.1em V}}\xspace}
\newcommand{\gev}{\ensuremath{\mathrm{\,Ge\kern -0.1em V}}\xspace}
\newcommand{\mev}{\ensuremath{\mathrm{\,Me\kern -0.1em V}}\xspace}
\newcommand{\kev}{\ensuremath{\mathrm{\,ke\kern -0.1em V}}\xspace}
\newcommand{\ev}{\ensuremath{\mathrm{\,e\kern -0.1em V}}\xspace}
\newcommand{\gevc}{\ensuremath{{\mathrm{\,Ge\kern -0.1em V\!/}c}}\xspace}
\newcommand{\mevc}{\ensuremath{{\mathrm{\,Me\kern -0.1em V\!/}c}}\xspace}
\newcommand{\gevcc}{\ensuremath{{\mathrm{\,Ge\kern -0.1em V\!/}c^2}}\xspace}
\newcommand{\gevgevcccc}{\ensuremath{{\mathrm{\,Ge\kern -0.1em V^2\!/}c^4}}\xspace}
\newcommand{\mevcc}{\ensuremath{{\mathrm{\,Me\kern -0.1em V\!/}c^2}}\xspace}

\def\mum  {\ensuremath{\,\upmu\rm m}\xspace}

\newcommand{\chisq}{\ensuremath{\chi^2}\xspace}

\def\gsim{{~\raise.15em\hbox{$>$}\kern-.85em
          \lower.35em\hbox{$\sim$}~}\xspace}
\def\lsim{{~\raise.15em\hbox{$<$}\kern-.85em
          \lower.35em\hbox{$\sim$}~}\xspace}

\def\pt         {\mbox{$p_{\rm T}$}\xspace}

\def\photos  {\mbox{\textsc{Photos}}\xspace}
\def\evtgen     {\mbox{\textsc{EvtGen}}\xspace}
\def\pythia     {\mbox{\textsc{Pythia}}\xspace}

\def\geant      {\mbox{\textsc{Geant4}}\xspace}

\def\tell1  {TELL1\xspace}
\def\ukl1   {UKL1\xspace}

\newcommand{\e}[1]{\times 10^{#1} }

\def\BtoDsphi  {\decay{B^{+}}{D_s^{+}\phi}}
\def\BtoDsDz   {\decay{\Bu}{\Dsp\Dzb}}

\def\DstoKKpi{\decay{D_s^+}{K^+K^-\pi^+}}

\def\DtoKpipi{\decay{D^+}{K^-\pi^+\pi^+}}

\def\ifb{fb$^{-1}$}

\usepackage{cite} 
\usepackage{mciteplus}

\begin{document}

\renewcommand{\thefootnote}{\fnsymbol{footnote}}
\setcounter{footnote}{1}


\begin{titlepage}
\pagenumbering{roman}

\vspace*{-1.5cm}
\centerline{\large EUROPEAN ORGANIZATION FOR NUCLEAR RESEARCH (CERN)}
\vspace*{1.5cm}
\hspace*{-0.5cm}
\begin{tabular*}{\linewidth}{lc@{\extracolsep{\fill}}r}
\ifthenelse{\boolean{pdflatex}}
{\vspace*{-2.7cm}\mbox{\!\!\!\includegraphics[width=.14\textwidth]{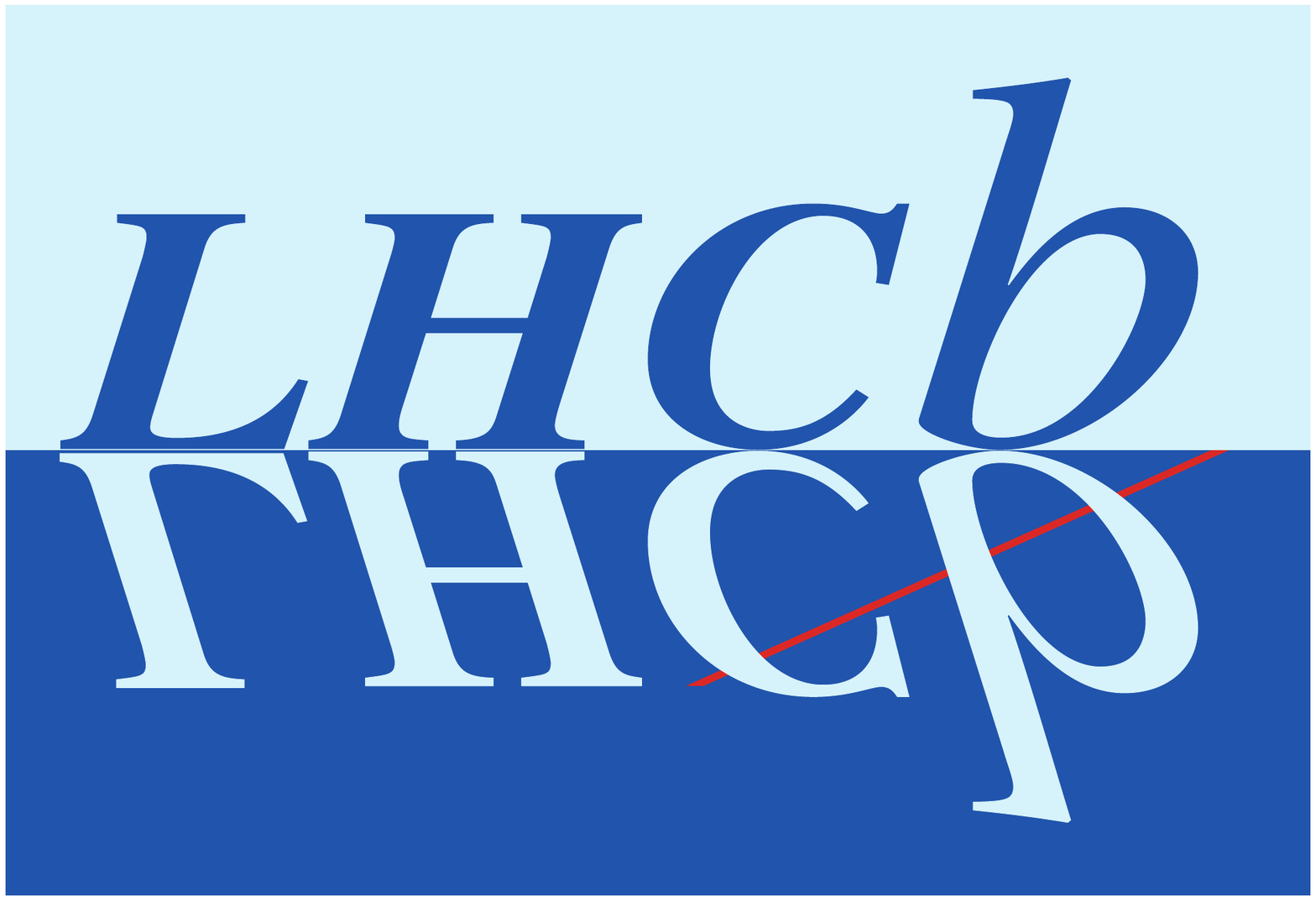}} & &}%
{\vspace*{-1.2cm}\mbox{\!\!\!\includegraphics[width=.12\textwidth]{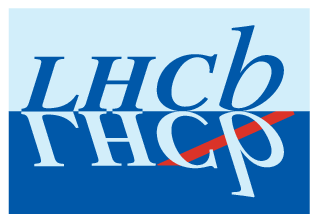}} & &}%
\\
 & & CERN-PH-EP-2012-286 \\  
 & & LHCb-PAPER-2012-025 \\  
 & & January 8, 2013 \\ 
 & & \\
\end{tabular*}

\vspace*{4.0cm}

{\bf\boldmath\huge
\begin{center}
  First evidence for the annihilation decay mode 
$B^{+} \to D_s^{+}\phi$
\end{center}
}

\vspace*{2.0cm}

\begin{center}
The LHCb collaboration\footnote{Authors are listed on the following pages.}
\end{center}

\vspace{\fill}

\begin{abstract}
  \noindent
Evidence for the hadronic annihilation decay mode $B^{+} \to D_s^{+}\phi$ is found
with greater than $3\sigma$ significance.
The branching fraction and \CP asymmetry are measured to be
\begin{eqnarray}
\mathcal{B}(B^{+} \to D_s^{+}\phi) &=& 
\left(1.87^{\,+1.25}_{\,-0.73}\,({\rm stat}) \pm 0.19\, ({\rm syst}) \pm 0.32\, ({\rm norm})\right) \times 10^{-6}, \nonumber\\
\mathcal{A}_{CP}(B^{+} \to D_s^{+}\phi) &=& -0.01 \pm 0.41\, ({\rm stat}) \pm 0.03\, ({\rm syst}).  \nonumber
\end{eqnarray}
The last uncertainty on $\mathcal{B}(B^{+} \to D_s^{+}\phi)$ is from the branching fractions of the \BtoDsDz normalization mode and intermediate resonance decays.   Upper limits are also set for the branching fractions of the related decay modes $B^{+}_{(c)} \to D^{+}_{(s)}\Kstarz$, ${B^{+}_{(c)} \to D^{+}_{(s)}\Kstarzb}$ and ${B_c^{+} \to D^{+}_{s}\phi}$, including the result   
${\mathcal{B}(B^+ \to \Dp\Kstarz)} < 1.8 \times 10^{-6}$ at the 90\% credibility level. 
\end{abstract}

\vspace*{0.0cm}

\begin{center}
Submitted to the Journal of High Energy Physics
\end{center}

\vspace{\fill}

\end{titlepage}


\newpage
\setcounter{page}{2}
\mbox{~}
\newpage

\centerline{\large\bf LHCb collaboration}
\begin{flushleft}
\small
R.~Aaij$^{38}$, 
C.~Abellan~Beteta$^{33,n}$, 
A.~Adametz$^{11}$, 
B.~Adeva$^{34}$, 
M.~Adinolfi$^{43}$, 
C.~Adrover$^{6}$, 
A.~Affolder$^{49}$, 
Z.~Ajaltouni$^{5}$, 
J.~Albrecht$^{35}$, 
F.~Alessio$^{35}$
% ...

{\footnotesize \it
% ...
$ ^{5}$Clermont Universit\'{e}, Universit\'{e} Blaise Pascal, CNRS/IN2P3, LPC, Clermont-Ferrand, France\\
$ ^{6}$CPPM, Aix-Marseille Universit\'{e}, CNRS/IN2P3, Marseille, France\\
% ...
$ ^{11}$Physikalisches Institut, Ruprecht-Karls-Universit\"{a}t Heidelberg, Heidelberg, Germany\\
% ...
$ ^{33}$Universitat de Barcelona, Barcelona, Spain\\
$ ^{34}$Universidad de Santiago de Compostela, Santiago de Compostela, Spain\\
$ ^{35}$European Organization for Nuclear Research (CERN), Geneva, Switzerland\\
% ...
$ ^{38}$Nikhef National Institute for Subatomic Physics, Amsterdam, The Netherlands\\
% ...
$ ^{43}$H.H. Wills Physics Laboratory, University of Bristol, Bristol, United Kingdom\\
% ...
$ ^{49}$Oliver Lodge Laboratory, University of Liverpool, Liverpool, United Kingdom\\
% ...
$ ^{n}$LIFAELS, La Salle, Universitat Ramon Llull, Barcelona, Spain\\
% ...
}
\end{flushleft}

\cleardoublepage


\renewcommand{\thefootnote}{\arabic{footnote}}
\setcounter{footnote}{0}



\pagestyle{plain} 
\setcounter{page}{1}
\pagenumbering{arabic}


\section{Introduction}
The decays\footnote{Throughout this paper, charge conjugation is implied.  Furthermore, $K^{*0}$ and $\phi$ denote the $K^{*0}(892)$ and $\phi(1020)$ resonances, respectively.} 
$B^+ \to D_s^+\phi,\; D^+ \Kstarz, \;D_s^+ \Kstarzb$
occur in the Standard Model (SM) via annihilation of the quarks forming the $B^+$ meson into a virtual $W^+$ boson 
(Fig.~\ref{fig:dsphi_diagram}).
There is currently strong interest in annihilation-type decays of $B^+$ mesons due, in part, to the roughly $2\sigma$ deviation above the SM prediction observed in the branching fraction of $B^+ \to \tau^+ \nu$~\cite{Lees:2012ju,ref:belletaunu}.   
Annihilation diagrams of $B^+$ mesons are highly suppressed in the SM; no hadronic annihilation-type decays of the $B^+$ meson have been observed to-date. 
Branching fraction predictions (neglecting rescattering) for \BtoDsphi and \decay{\Bp}{\Dp\Kstarz} are $(1-7) \times 10^{-7}$ in the SM~\cite{Zou:2009zza,Mohanta:2002wf,PhysRevD.76.057701,Lu:2001yz}, where the precision of the calculations is limited by hadronic uncertainties.  The branching fraction for the \decay{\Bp}{\Dsp\Kstarzb} decay mode is expected to be about 20 times smaller
due to the CKM quark-mixing matrix
elements involved.  
The current upper limits on the branching fractions of these decay modes are $\mathcal{B}(\BtoDsphi) < 1.9 \times 10^{-6}$~\cite{Aubert:2005gd},  
$\BR(\decay{\Bp}{\Dp\Kstarz}) < 3.0\e{-6}$~\cite{delAmoSanchez:2010rf}
and $\BR(\decay{\Bp}{\Dsp\Kstarzb}) < 4.0\e{-4}$~\cite{Alexander:1993gp}, 
all at the 90\% confidence level.

Contributions from physics beyond the SM (BSM) could greatly enhance these branching fractions and/or produce a large \CP asymmetry~\cite{Mohanta:2002wf,PhysRevD.76.057701}.
For example, a charged Higgs ($H^+$) boson mediates the annihilation process.  
Interference between the $W^+$ and $H^+$ amplitudes could result in a \CP asymmetry if the two amplitudes are of comparable size and have both strong and weak phase differences different from zero.
An $H^+$ contribution to the amplitude could also significantly increase the branching fraction.

In this paper, first evidence for the decay mode \BtoDsphi is presented using 1.0~\ifb\ of data collected by \lhcb in 2011 from $pp$ collisions at a center-of-mass energy of 7~TeV.  The branching fraction and \CP asymmetry are measured.  Limits are set on the branching fraction of the decay modes \decay{\Bp}{\Dp\Kstarz} and \decay{\Bp}{\Dsp\Kstarzb}, along with the highly suppressed decay modes \decay{\Bp}{\Dp\Kstarzb} and \decay{\Bp}{\Dsp\Kstarz}.   Limits are also set on the product of the production rate and branching fraction for $B_c^+$ decays to the final states $D_s^+\phi$, $D_{(s)}^+K^{*0}$ and $D_{(s)}^+ \Kstarzb$.

\begin{figure} 
\centering
\includegraphics[width=0.4\textwidth]{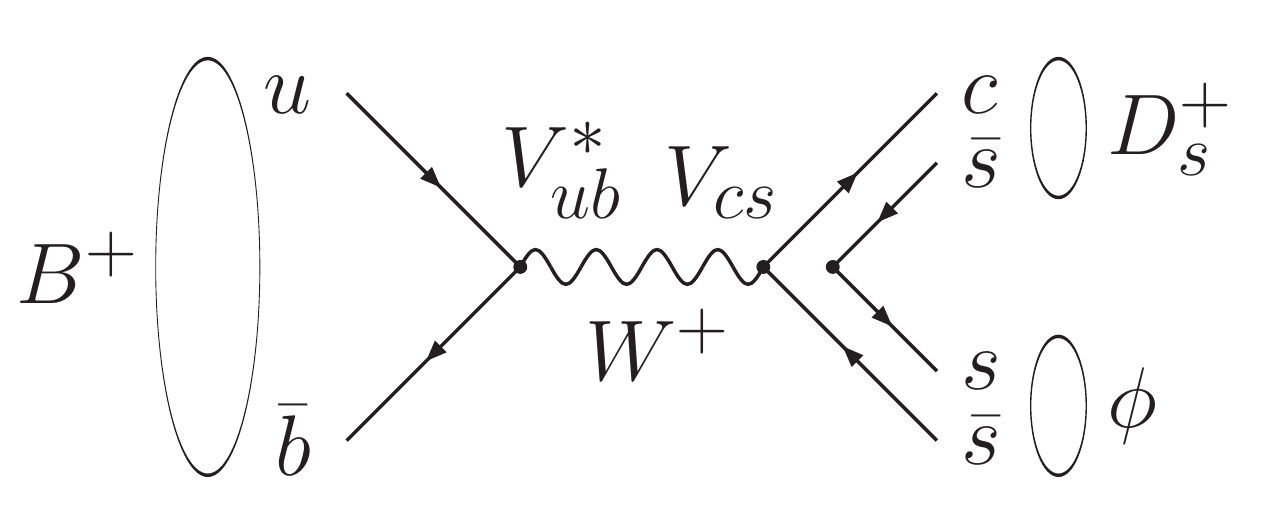}\\
\includegraphics[width=0.4\textwidth]{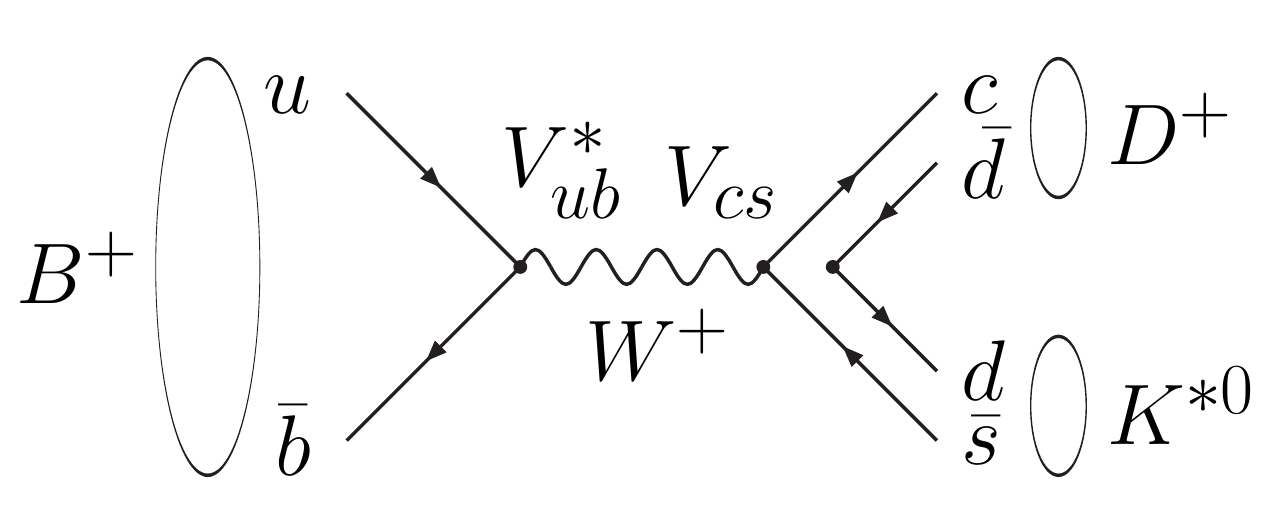} 
\includegraphics[width=0.4\textwidth]{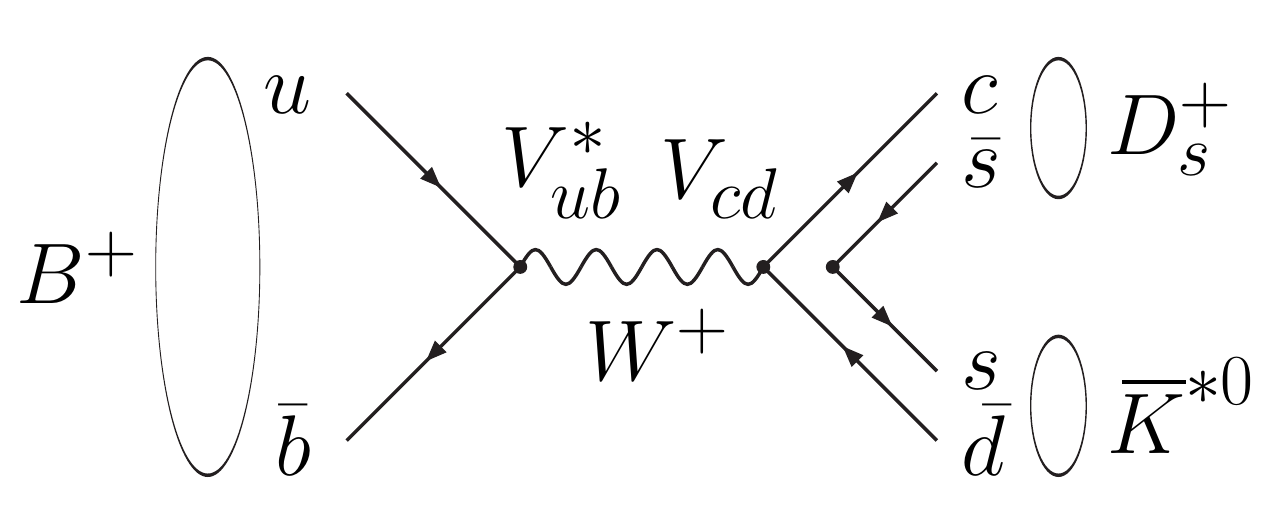}
\caption{
\label{fig:dsphi_diagram}
Feynman diagrams for \BtoDsphi, \decay{\Bp}{\Dp\Kstarz} and \decay{\Bp}{\Ds\Kstarzb} decays. 
}
\end{figure}

\section{The \lhcb experiment}

The \lhcb detector~\cite{Alves:2008zz} is a single-arm forward
spectrometer covering the \mbox{pseudorapidity} range $2<\eta <5$, designed
for the study of particles containing \bquark or \cquark quarks. The
detector includes a high precision tracking system consisting of a
silicon-strip vertex detector surrounding the $pp$ interaction region,
a large-area silicon-strip detector located upstream of a dipole
magnet with a bending power of about $4{\rm\,Tm}$, and three stations
of silicon-strip detectors and straw drift tubes placed
downstream. The combined tracking system has a momentum resolution
$\Delta p/p$ that varies from 0.4\% at 5\gevc to 0.6\% at 100\gevc,
and an impact parameter resolution of 20\mum for tracks with high
transverse momentum (\pt).  Discrimination between different types of charged particles is provided by two ring-imaging Cherenkov detectors~\cite{LHCb:RICH}. Photon, electron and hadron
candidates are identified by a calorimeter system consisting of
scintillating-pad and preshower detectors, an electromagnetic
calorimeter and a hadronic calorimeter. Muons are identified by a muon
system composed of alternating layers of iron and multiwire
proportional chambers. 

The \lhcb trigger~\cite{LHCb:TRIG} consists of a hardware stage, based
on information from the calorimeter and muon systems, followed by a
software stage which applies a partial event reconstruction (only tracks with $\pt > 0.5$\gevc are used).
The software stage of the \lhcb trigger builds two-, three- and four-track partial $b$-hadron candidates that are required to be significantly displaced from the primary interaction and have a large sum of $\pt$ in their tracks.  At least one of the tracks used to form the trigger candidate must have $\pt > 1.7$\gevc and impact parameter  \chisq with respect to the primary interaction $\chi^2_{\rm IP} > 16$.   The $\chi^2_{\rm IP}$ is defined as the difference between the $\chi^2$ of the primary interaction vertex reconstructed with and without the considered track.
A boosted decision tree (BDT)~\cite{ref:bdt,ref:bdt2} is used to distinguish between trigger candidates originating from $b$-hadron decays and those that originate from prompt $c$-hadrons or combinatorial background.  The BDT 
provides a pure sample of $b\bar{b}$ events for offline analysis.

For the simulation, $pp$ collisions are generated using
\pythia~6.4~\cite{Sjostrand:2006za} with a specific \lhcb
configuration~\cite{LHCb-PROC-2010-056}.  Decays of hadronic particles
are described by \evtgen~\cite{Lange:2001uf} in which final state
radiation is generated using \photos~\cite{Golonka:2005pn}. The
interaction of the generated particles with the detector and its
response are implemented using the \geant
toolkit~\cite{Allison:2006ve, *Agostinelli:2002hh} as described in
Ref.~\cite{LHCb-PROC-2011-006}.

\section{Event selection}
\label{sec:sel}

Candidates of the decays searched for are formed from tracks that are required to have $\pt > 0.1$\gevc, $\chi^2_{\rm IP} > 4$ and $p > 1$\gevc.   For the $\phi$ and $K^{*0}$ decay products the momentum requirement is increased to $p > 2$\gevc.  These momentum requirements are 100\% efficient on simulated signal events. 
The $D_s^+ \to K^+K^-\pi^+$, ${D^+ \to K^-\pi^+\pi^+}$, ${\phi\to K^+K^-}$ and $K^{*0}\to K^+\pi^-$ candidates are required to have invariant masses within 25, 25, 20 and 50\mevcc of their respective world-average (PDG) values~\cite{ref:pdg}.   The mass resolutions for  $D_s^+ \to K^+K^-\pi^+$ and ${D^+ \to K^-\pi^+\pi^+}$ are about 7\mevcc and 8\mevcc, respectively.
The decay chain is fit constraining the $D^+_{(s)}$ candidate mass to its PDG value.  The $D^+_{(s)}$ vertex is required to be downstream of the $B^+$ vertex and the $p$-value formed from $\chi^2_{\rm IP}+\chi^2_{\rm vertex}$ of the $B^+$ candidate is required to be greater than 0.1\%.  Backgrounds from charmless decays are suppressed by requiring significant separation between the $D^+_{(s)}$ and $B^+$ decay vertices.  This requirement reduces contributions from charmless backgrounds by a factor of about 15 while retaining 87\% of the signal.

Cross-feed between $D^+$ and $D_s^+$ candidates can occur if one of the child tracks is misidentified.  If a \DstoKKpi candidate can also form a $D^+ \to K^-\pi^+\pi^+$ candidate that falls within 25\mevcc of the PDG $D^+$ mass, then it is rejected unless either ${|m_{KK}-m_{\phi}^{\rm PDG}| < 10}$\mevcc or the ambiguous child track satisfies a stringent kaon particle identification (PID) requirement.   This reduces the $D^+ \to D_s^+$ cross-feed by a factor of about 200 at the expense of only 4\% of the signal.  
For decay modes that contain a $D^+$ meson, 
a \DtoKpipi candidate that can also form a $D^+_s \to K^-K^+\pi^+$ candidate whose mass is within 25\mevcc of the PDG $D^+_s$ mass is rejected if either ${|m_{KK}-m_{\phi}^{\rm PDG}| < 10}$\mevcc or the ambiguous child track fails a stringent pion PID requirement.  For all modes, $\Lambda_c^+ \to D^+_{(s)}$ cross-feed (from the $\Lambda_c^+ \to pK^-\pi^+$ decay mode) is suppressed using similar requirements.

When a pseudoscalar particle decays into a pseudoscalar and a vector, $V$, the spin of the vector particle (in this case a \Pphi or \Kstarz) must be orthogonal to its momentum to conserve angular momentum; {\em i.e.}, the vector particle must be longitudinally polarized.
For a longitudinally-polarized \Pphi(\Kstarz) decaying into the $K^+K^-(K^+\pi^-$) final state, the angular distribution
of the $K^+$ meson in the $V$ rest frame is proportional to $\cos^2{\theta_{K}}$, where
$\theta_K$ is the angle between the momenta of the \Kp and \Bp in the $V$ rest frame.
The requirement $|\cos{\theta_K}| > 0.4$, which is 93\% efficient on signal and rejects about 40\% of the background, is applied in this analysis.  

Four BDTs that identify \DstoKKpi, ${D^+ \to K^-\pi^+\pi^+}$, $\phi \to K^+K^-$ and ${K^{*0} \to K^+\pi^-}$ candidates originating from $b$-hadron decays are used to suppress the backgrounds.   The BDTs are trained using large clean $D_{(s)}^+$, $\phi$ and $K^{*0}$ samples obtained from $\Bb^0_{(s)} \to D_{(s)}^+\pi^-$,  $B_s^0 \to \jpsi\phi$ and $B^0 \to \jpsi K^{*0}$ data, respectively, where the backgrounds are subtracted using the sPlot technique~\cite{2005NIMPA.555..356P}.  Background samples for the training are taken from the $D_{(s)}^+$, $\phi$ and $K^{*0}$ sidebands in the same data samples.
The BDTs take advantage of the kinematic similarity of all $b$-hadron decays and avoid using any topology-dependent information.
The BDTs use kinematic, track quality, vertex and PID information to obtain a high level of background suppression.   
In total, 23 properties per child track and five properties from the parent $D_{(s)}^+$, \Pphi or \Kstarz meson are used in each BDT.   The boosting method used is known as {\em bagging}~\cite{ref:bag}, which produces BDT response values
in the unit interval.

A requirement is made on the product of the BDT responses of the $D_{(s)}^+$ and $\phi$ or $K^{*0}$ candidates. Tests on several $B^{0}_{(s)} \to DD^{\prime}$ decay modes show that this provides the best performance~\cite{ref:b2dd}.  The efficiencies of these cuts are obtained using large 
$\Bb_{(s)}^0 \to D_{(s)}^+\pi^-$,  $B_s^0 \to \jpsi\phi$ and $B^0 \to \jpsi K^{*0}$
data samples that are not used in the BDT training.
The efficiency calculation takes into account the kinematic differences between the signal and training decay modes using additional input from simulated data.  Correlations between the properties of the $D_{(s)}^+$ and \Pphi or \Kstarz mesons in a given $B^+$ candidate are also accounted for.  

The optimal BDT requirements are chosen such that the signal significance is maximized for the central value of the available SM branching fraction predictions.
The signal efficiency of the optimal BDT requirement is 51\%, 69\% and 51\% for $B^+ \to D^+_s\phi$, $B^+ \to D^+ K^{*0}$ and $B^+ \to D_s^+ \Kstarzb$ decay modes, respectively.  
The final sample contains no events with multiple candidates. 
Finally, no consideration is given to contributions where the $K^+K^-(K^+\pi^-)$ is in an $S$-wave state or from the tails of higher $\phi(K^{*0})$ resonances.
Such contributions are neglected as they are expected to be much smaller than the  statistical uncertainties.

\boldmath
\section{Branching fraction for the $B^+ \to D_s^+ \phi$ decay}
\label{sec:sig-yield}
\unboldmath

\begin{table}
  \caption{\label{tab:fitregions} Summary of fit regions for $\Bp \to \Dsp\phi$. About 89\% of the signal is expected to be in region A.}
  \begin{center}
    \begin{tabular}{c|cc}
    {} & \multicolumn{2}{c}{$|m_{K\kern-0.1em{K}}-m_\phi|$ (\mevcc)} \\
    $|\cos\theta_K|$ & $< 20$ & $(20,40)$ \\ 
      \hline
      $> 0.4$ &  A & B       \\ 
      $< 0.4$ & C & D   \\ 
    \end{tabular}
  \end{center}
\end{table}

The \BtoDsphi yield is determined by performing an unbinned maximum likelihood fit to the invariant mass spectra of \Bp candidates.  
Candidates failing the $\cos{\theta_K}$ and/or $m_{KK}$ selection criteria that are within 40\mevcc of $m_{\phi}^{\rm PDG}$ are used in the fit to help constrain the background probability density function (PDF).
The data set is comprised of the four subsamples given in Table~\ref{tab:fitregions}.
They are fit simultaneously to a PDF with the following components:
\begin{itemize}
  \item \BtoDsphi: A Gaussian function whose parameters are taken from simulated data and fixed in the fit is used for the signal shape.
The fraction of signal events in each of the subsamples is also fixed from simulation to be as follows: (A) 89\%; (B) 4\%; (C) 7\% and (D) no signal expected.  
Thus, almost all signal events are expected to be found in region A, while region D should contain only background.
A 5\% systematic uncertainty is assigned to the branching fraction determination due to the shape of the signal PDF.  This value is obtained by considering the effect on the branching fraction for many variations of the signal PDFs for \BtoDsphi and the normalization decay mode.
  \item $B^{+} \to D_s^{*+} \phi$:  The $\phi$ in this decay mode does not need to be longitudinally polarized.  When the photon from the $D_s^{*+}$ decay is not reconstructed, the polarization affects both the invariant mass distribution and the fraction of events in each of the subsamples.  Studies using a wide range of polarization fractions, with shapes taken from simulation, show that the uncertainties in this PDF have a negligible impact on the signal yield.  
\item $\Bsb \to D_s^{(*)+}K^-\Kstarz$:  These decay modes, which arise as backgrounds to $\BtoDsphi$ when the pion from the \Kstarz decay is not reconstructed, have not yet been observed; however, they are expected to have similar branching fractions to the decay modes $\Bzb \to D^{(*)+}K^-\Kstarz$.  The ratio ${\mathcal{B}(\Bsb \to D_s^{*+}K^-\Kstarz)}/{\mathcal{B}(\Bsb \to D_s^+K^-\Kstarz)}$ is fixed to be the same as the value of $\mathcal{B}(\Bzb \to D^{*+}K^-\Kstarz)/{\mathcal{B}(\Bzb \to D^+K^-\Kstarz)}$~\cite{ref:belledkkst}.  The fraction of events in each subsample is constrained by simulation.  Removing these constraints results in a 1\% change in the signal yield.
\item Combinatorial background: An exponential shape is used for this component.  The exponent is fixed to be the same in all four subsamples.  This component is assumed to be uniformly distributed in $\cos{\theta_K}$.  Removing these constraints produces shifts in the signal yield of up to 5\%; thus, a 5\% systematic uncertainty is assigned to the branching fraction measurement.
\end{itemize}
To summarize, the parameters allowed to vary in the fit are the signal yield, the yield and longitudinal polarization fraction of $B^{+} \to D_s^{*+} \phi$, the yield of $\Bsb \to D_s^{(*)+}K^-\Kstarz$ in each subsample, the combinatorial background yield in each subsample and the combinatorial exponent.

Figure~\ref{fig:fit-full} shows the $B^+$ candidate invariant mass spectra for each of the four subsamples, along with the various components of the PDF.  The signal yield is found to be $6.7^{\,+4.5}_{\,-2.6}$, where the confidence interval includes all values of the signal yield for which $\log{(\mathcal{L}_{\rm max}/\mathcal{L})} < 0.5$.  The statistical significance of the signal is found using Wilks Theorem~\cite{ref:wilks}
to be $3.6\sigma$.  A simulation study consisting of an ensemble of $10^5$ data sets confirms the significance and also the accuracy of the coverage to within a few percent.  All of the variations in the PDFs discussed above result in significances above $3\sigma$; thus, evidence for 
\BtoDsphi
is found at greater than $3\sigma$ significance including systematics.    

\begin{figure} 
\centering
    \includegraphics[width=0.99\textwidth]{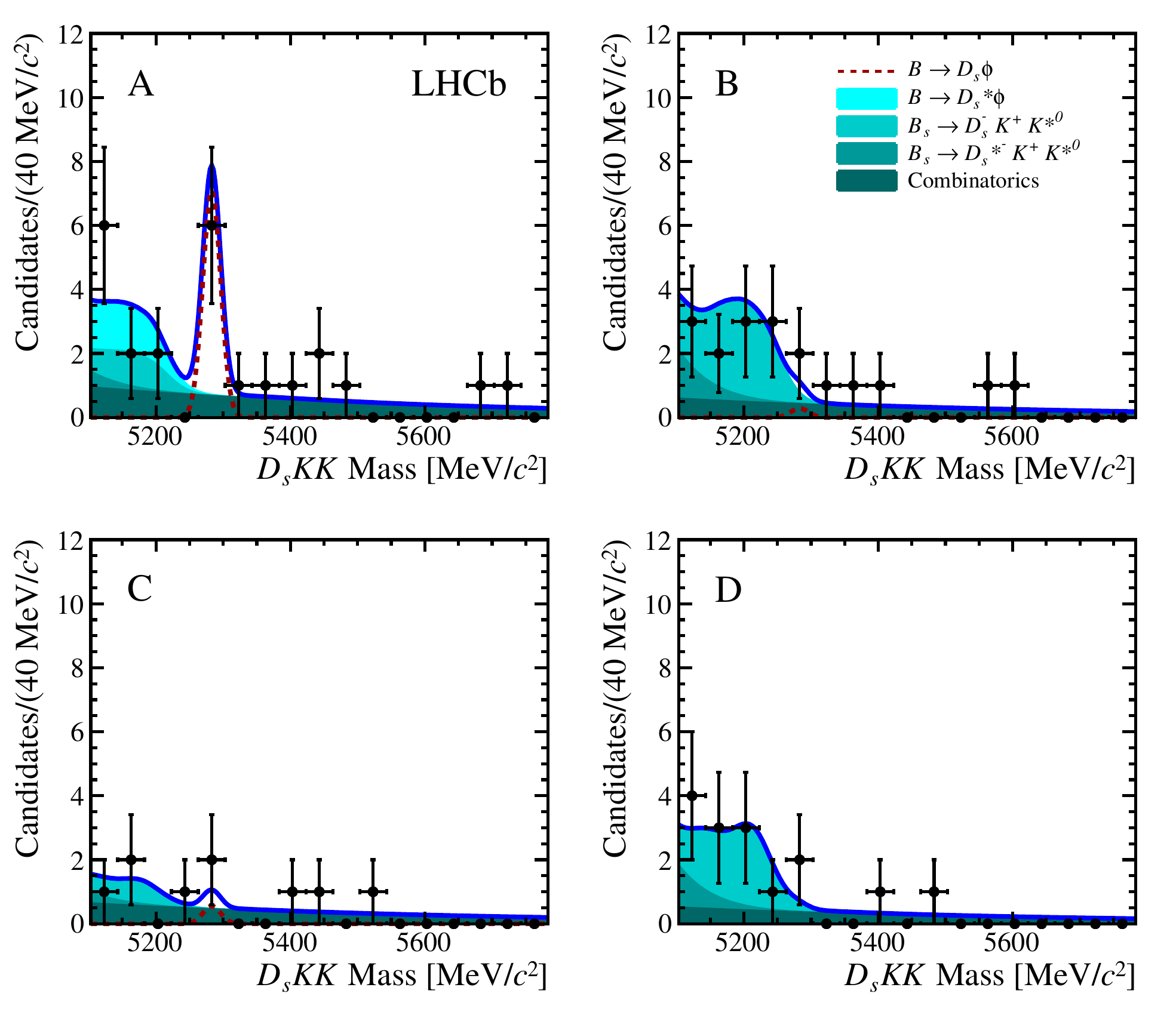}
\caption{
\label{fig:fit-full}
Fit results for \BtoDsphi.  The fit regions, as given in Table~\ref{tab:fitregions}, are labelled on the panels. 
The PDF components are as given in the legend.
}
\end{figure}

The \BtoDsphi branching fraction is normalized to $\mathcal{B}(\BtoDsDz)$. The selection for the normalization mode, which is similar to that used here for \BtoDsphi, is described in detail in Ref.~\cite{ref:b2dd}.  The ratio of the efficiency of the product of the geometric, trigger, reconstruction and selection (excluding the charmless background suppression and BDT) requirements of the signal mode to the normalization mode is found from simulation to be $0.93\pm0.05$.  The ratio of BDT efficiencies, which include all usage of PID information, is determined from data (see Sect.~\ref{sec:sel}) to be
$0.52\pm0.02$.  The large branching fraction of the normalization mode permits using a BDT requirement that is nearly 100\% efficient.
For the charmless background suppression requirement, 
the efficiency ratio is determined from simulation to be $1.15\pm0.01$.  The difference is mostly due to the fact that the normalization mode has two charmed mesons, while the signal mode only has one. 
The branching fraction is measured as
\begin{eqnarray}
  \mathcal{B}(\BtoDsphi) &=& \frac{\epsilon(\BtoDsDz)}{\epsilon(\BtoDsphi)}
        \frac{\mathcal{B}(\Dzb \to K^-\pi^+)}{\mathcal{B}(\phi \to K^+K^-)} \frac{N(\BtoDsphi)}{N(\BtoDsDz)} \,\mathcal{B}(\BtoDsDz) \nonumber \\ 
  {} &=& \left(1.87^{\,+1.25}_{\,-0.73}\,({\rm stat}) \pm 0.19\, ({\rm syst}) \pm 0.32\, ({\rm norm})\right) \times 10^{-6} \nonumber,
\end{eqnarray}  
where $\epsilon$ denotes efficiency.  The normalization uncertainty includes contributions from $\mathcal{B}(\BtoDsDz) = (1.0\pm0.17)\%$, $\mathcal{B}(\Dzb \to K^-\pi^+) = (3.88 \pm 0.05)\%$ and ${\mathcal{B}(\phi \to K^+K^-)} = {(48.9 \pm 0.5)\%}$~\cite{ref:pdg}.  The systematic uncertainties are summarized in Table~\ref{tab:sys}.  
The value obtained for $\mathcal{B}(\BtoDsphi)$ is consistent with the SM calculations given the large uncertainties on both the theoretical and experimental values.  

\begin{table}[t]
  \caption{Systematic uncertainties contributing to $\mathcal{B}(\BtoDsphi)/\mathcal{B}(\BtoDsDz)$.}
\begin{center}\begin{tabular}{l|c}
    Source & Uncertainty (\%) \\
    \hline
    Selection      & $\phantom{1}7$\\
    Signal PDF     & $\phantom{1}5$\\
    Background PDF & $\phantom{1}5$\\
    Normalization & 17 \\
\end{tabular}\end{center}
\label{tab:sys}
\end{table}

\boldmath
\section{Branching fractions for the decays ${B^{+} \to D_{(s)}^{+} \Kstarz}$ and ${B^+ \to D_{(s)}^{+}\Kstarzb}$}
\label{sec:limits}
\unboldmath

The SM predicts the branching fraction ratios $\mathcal{B}(B^+\to\Dp\Kstarz)/\mathcal{B}(\B^+\to D_s^+\phi) \sim 1$ and ${\mathcal{B}(B^+\to\Dsp\Kstarzb)}/{\mathcal{B}(B^+\to D_s^+\phi)} \sim |V_{cd}/V_{cs}|^2$~\cite{Zou:2009zza}.
The partially reconstructed backgrounds are expected to be much larger in these channels compared to \BtoDsphi mainly due to the large \Kstarz mass window.  Producing an exhaustive list of decay modes that contribute to each of these backgrounds is not feasible; thus, reliable PDFs for the backgrounds are not available. 
Instead, data in the sidebands around the signal region are used to estimate the expected background yield in the signal region.   The signal region is chosen to be $\pm2\sigma$ around the $B^+$ mass, where $\sigma = 13.8$\mevcc is determined from simulation.   

Our prior knowledge about the background can be stated as the following three assumptions: (1) the slope is negative, which will be true provided $b$-baryon background contributions are not too large; 
(2) it does not peak or form a shoulder\footnote{No evidence of peaking backgrounds is found in either the $D_{(s)}^+$ or \Kstarz sidebands. If peaking backgrounds do make significant contributions, then the limits set in this paper are conservative.} 
and (3) the background yield is non-negative.  These background properties are assumed to hold throughout the signal and sideband regions. 
To convert these assumptions into background expectations, ensembles of background-only data sets are generated using the observed data in the sidebands and assuming Poisson distributed yields.  For each simulated data set, all interpolations into the signal region that satisfy our prior assumptions are assigned equal probability.  These probabilities are summed over all data sets to produce background yield PDFs, all of which are well described by Gaussian lineshapes (truncated at zero) with the parameters $\mu_{\rm bkgd}$ and $\sigma_{\rm bkgd}$ given in Table~\ref{tab:kstar_lims}.  
The $B^+$ candidate invariant mass 
distributions, along with the background expectations, are shown in Fig.~\ref{fig:bu_kstar}.  
The results of spline interpolation using data in the sideband bins, along with the 68\% confidence intervals obtained by propagating the Poisson uncertainties in the sidebands to the splines, are shown for comparison.
As expected, the spline interpolation results, which involve a stronger set of assumptions, have less statistical uncertainty.  

\begin{figure} 
\centering
    \includegraphics[width=0.45\textwidth]{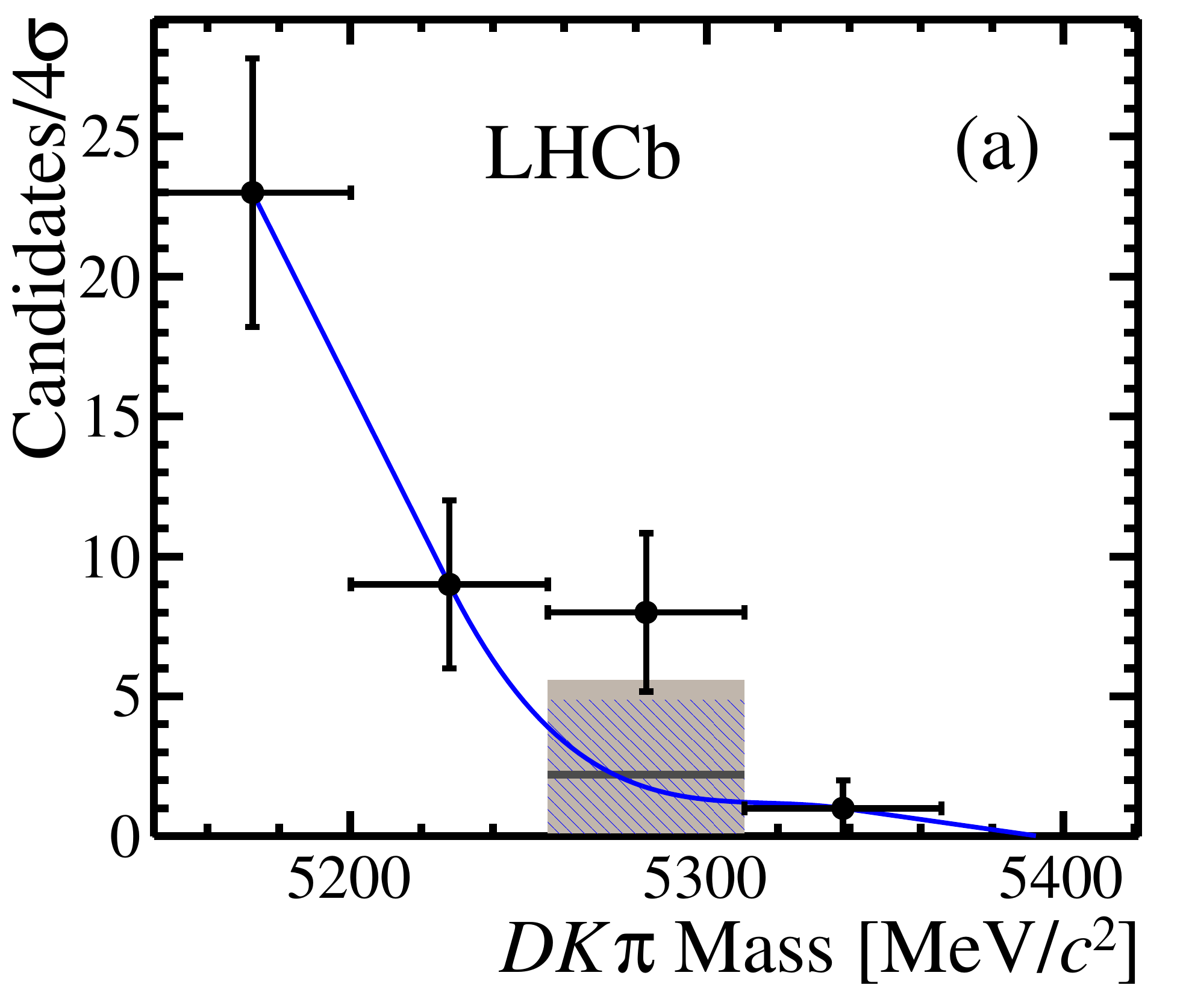}
    \includegraphics[width=0.45\textwidth]{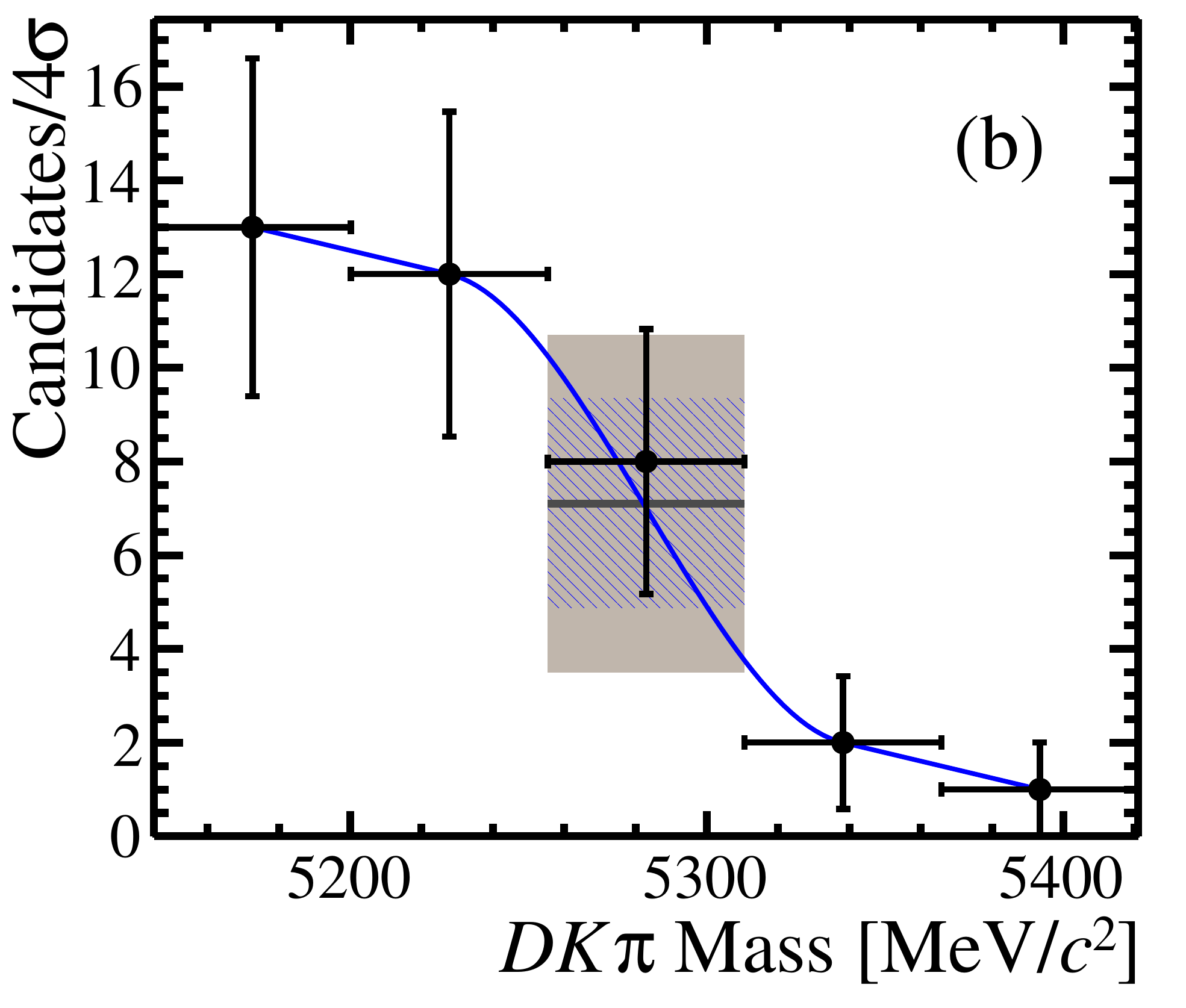}\\
    \includegraphics[width=0.45\textwidth]{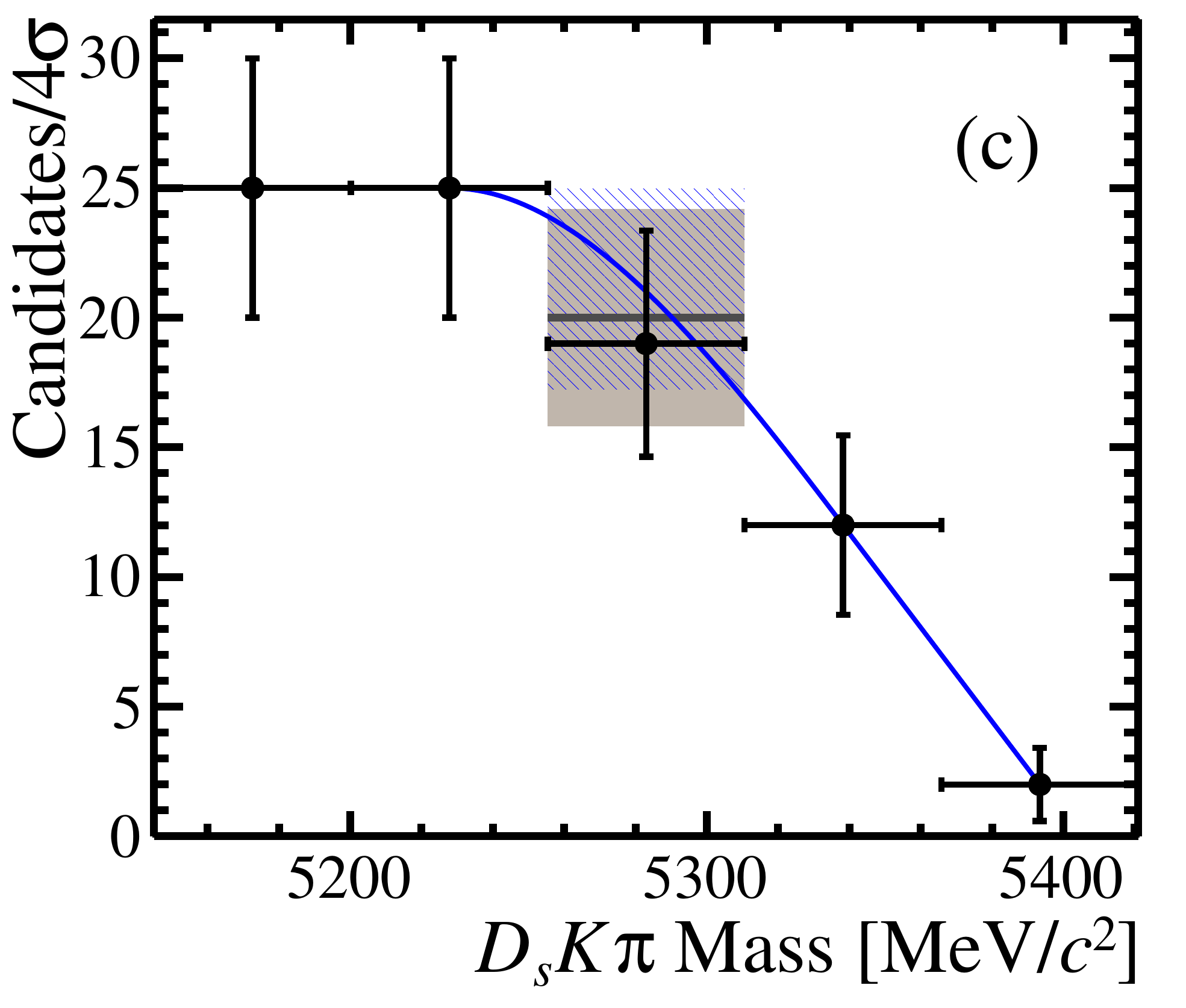}
    \includegraphics[width=0.45\textwidth]{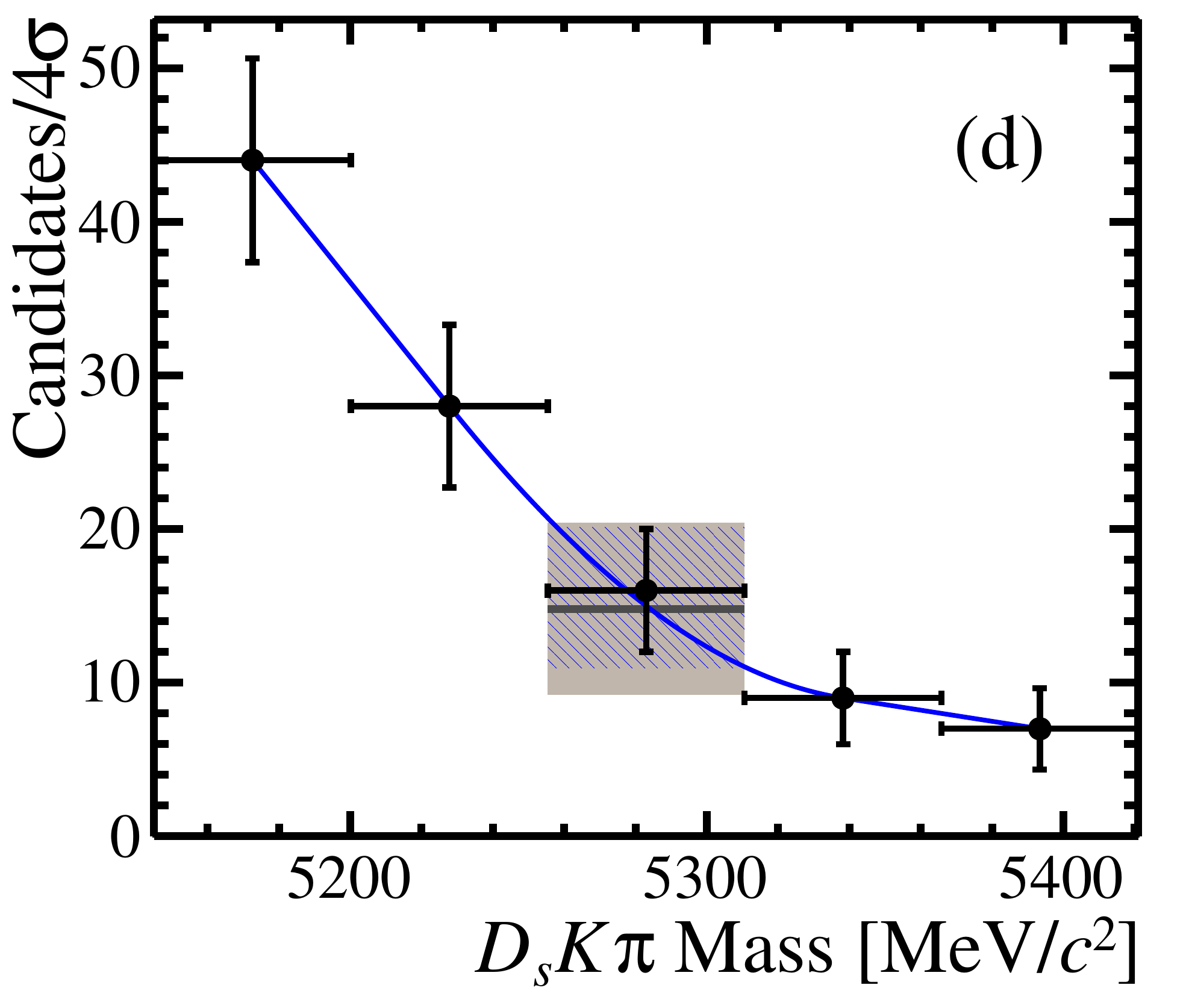}
\caption{
\label{fig:bu_kstar}
Invariant mass distributions for 
(a)~\decay{\Bp}{\Dp\Kstarz},
(b)~\decay{\Bp}{\Dp\Kstarzb},
(c)~${\Bp \to \Dsp\Kstarz}$
and    
(d)~\decay{\Bp}{\Dsp\Kstarzb}. The bins are each $4\sigma$ wide, where ${\sigma =13.8}$~\mevcc is the expected width of the signal peaks (the middle bin is centred at the expected $B^+$ mass).
The shaded regions are the $\mu_{\rm bkgd}\pm\sigma_{\rm bkgd}$ intervals (see Table~\ref{tab:kstar_lims}) used for the limit calculations; they are taken from the truncated-Gaussian priors as discussed in the text.
Spline interpolation results (solid blue line and hashed blue areas) are shown for comparison.
}
\end{figure}

A Bayesian approach~\cite{ref:bays} is used to set the upper limits.  Poisson distributions are assumed for the observed candidate counts and uniform, non-negative prior PDFs for the signal branching fractions.  The systematic uncertainties in the efficiency and {\BtoDsDz} normalization are encoded in log-normal priors, while the background prior PDFs are the truncated Gaussian lineshapes discussed above.  The posterior PDF, $p(\mathcal{B}|n_{\rm obs})$, where $n_{\rm obs}$ is the number of candidates observed in the signal region, is computed by integrating over the background, efficiency and normalization.  The 90\% credibility level (CL) upper limit, $\mathcal{B}_{90}$, is the value of the branching fraction for which $\int_0^{\mathcal{B}_{90}} p(\mathcal{B}|n_{\rm obs}) {\rm d}\mathcal{B} = 0.9 \int_0^{\infty}p(\mathcal{B}|n_{\rm obs}) {\rm d}\mathcal{B}$.  The upper limits are given in Table~\ref{tab:kstar_lims}.  The limit on \decay{\Bp}{\Dp\Kstarz} is 1.7 times lower than any previous limit, while the \decay{\Bp}{\Ds\Kstarzb} limit is 91 times lower.  For the highly suppressed decay modes \decay{\Bp}{\Dp\Kstarzb} and \decay{\Bp}{\Dsp\Kstarz} these are the first limits to be set. 

\begin{table}[t]
  \caption{Upper limits on $\BF(B^\pm \to D_{(s)}^{\pm} \Kstarz)$, where $n_{\rm obs}$ is the number of events observed in each of the signal regions, while $\mu_{\rm bkgd}$ and $\sigma_{\rm bkgd}$ are the Gaussian parameters used in the background prior PDFs.}
\begin{center}\begin{tabular}{l|ccc|c}
    Decay & $n_{\rm obs}$ & $\mu_\mathrm{bkgd}$ & $\sigma_\mathrm{bkgd}$ & Upper Limit at 90\% CL \\
    \hline
    \decay{\Bp}{\Dp\Kstarz} & \phantom{0}8 & \phantom{0}2.2 & 3.4 & $1.8 \times 10^{-6}$  \\
    \decay{\Bp}{\Dp\Kstarzb} & \phantom{0}8 & \phantom{0}7.1 & 3.6 & $1.4 \times 10^{-6}$ \\
    \decay{\Bp}{\Ds\Kstarz} & 19 & 20.0 & 4.2 & $3.5 \times 10^{-6}$ \\
    \decay{\Bp}{\Ds\Kstarzb} & 16 & 14.8 & 5.6 & $4.4 \times 10^{-6}$ \\
\end{tabular}\end{center}
\label{tab:kstar_lims}
\end{table}

The posterior PDF for the \decay{\Bp}{\Dp\Kstarz} decay excludes the no-signal hypothesis at the 89\% CL 
and gives a  branching fraction measurement of $\mathcal{B}(\decay{\Bp}{\Dp\Kstarz})= (0.8^{\,+0.6}_{\,-0.5}) \times 10^{-6}$, where the uncertainty includes statistics and systematics.  This result is consistent with both the SM expectation
and, within the large uncertainties, with the value obtained above for $\mathcal{B}(\BtoDsphi)$. 
If processes beyond the SM are producing an enhancement in $\mathcal{B}(\BtoDsphi)$, then a similar effect would also be expected in \decay{\Bp}{\Dp\Kstarz}.  While an enhancement cannot be ruled out by the data, the combined $\mathcal{B}(\BtoDsphi)$ and $\mathcal{B}(\decay{\Bp}{\Dp\Kstarz})$ result is consistent with the SM interpretation.

\boldmath
\section{Limits on branching fractions of $B_c^+$ decay modes}
\label{sec:bc_limits}
\unboldmath

Annihilation amplitudes are expected to be much larger for $B_c^+$ decays due to the large ratio of $|V_{cb}/V_{ub}|$.  In addition, the $B_c^+ \to D_s^+ \phi,\; D^+ \Kstarz, \; D_s^+ \Kstarzb$ decay modes can also proceed via penguin-type diagrams. However, due to the fact that $B_c^+$ mesons are produced much more rarely than $B^+$ mesons in 7\tev $pp$ collisions (the ratio of $B_c^+$ to $B^+$ mesons produced is denoted by $f_c/f_u$), no signal events are expected to be observed in any of these $B_c^+$ channels.  
The Bayesian approach is again used to set the limits.  
A different choice is made here for the background prior PDFs because the background levels are so low.
The background prior PDFs are now taken to be Poisson distributions, where the observed background counts are obtained using regions of equal size to the signal regions in the high-mass sidebands.  Only the high-mass sidebands are used to avoid possible contamination from partially reconstructed $B_c^+$ backgrounds.  
In none of the decay modes is more than a single candidate seen across the combined signal and background regions. 
The limits obtained, which are set on the product of $f_c/f_u$ and the branching fractions (see Table~\ref{tab:bc_limits}),
are four orders of magnitude better than any previous limit set for a $B_c^+$ decay mode that does not contain charmonium.
As expected given the small numbers of candidates observed, the limits have some dependence on the choice made for the signal prior PDF.   As a cross check, the limits were also computed using various frequentist methods.   The largest difference found is 20\%.

\begin{table}[t]
  \caption{Upper limits on $f_c/f_u \cdot \mathcal{B}(B_c \to X)$, where $n_{\rm obs}$ and $n_{\rm bkgd}$ are the number of events observed in the signal and background (sideband) regions, respectively.}
\begin{center}\begin{tabular}{l|cc|c}
    Decay & $n_{\rm obs}$ & $n_{\rm bkgd}$ & Upper Limit at 90\% CL \\
    \hline
    $B_c^+ \to D_s^+\phi$ & 0 & 0 & $0.8 \times 10^{-6}$ \\
    $B_c^+ \to \Dp\Kstarz$ & 1 & 0 & $0.5 \times 10^{-6}$ \\
    $B_c^+ \to \Dp\Kstarzb$ & 0 & 0 & $0.4 \times 10^{-6}$ \\
    $B_c^+ \to \Dsp\Kstarz$ & 0 & 0 & $0.7 \times 10^{-6}$ \\
    $B_c^+ \to \Dsp\Kstarzb$ & 1 & 0 & $1.1 \times 10^{-6}$ \\
\end{tabular}\end{center}
\label{tab:bc_limits}
\end{table}

\boldmath
\section{\CP asymmetry for the decay \BtoDsphi}
\label{sec:cpv}
\unboldmath

To measure the \CP asymmetry, $\mathcal{A}_{CP}$, in \BtoDsphi, only candidates in region (a) and in a $\pm2\sigma$ window ($\pm 26.4$\mevcc) around the $B^+$ mass are considered.  The number of $B^+$ candidates is $n_+ = 3$, while the number of $B^-$ candidates is $n_- = 3$.  The integral of the background PDF from the fit described in detail in Sect.~\ref{sec:sig-yield} in the signal region is $n_{\rm bkgd} = 0.75$ (the background is assumed to be charge symmetric).  The observed charge asymmetry is $\mathcal{A}_{\rm obs} = (n_- - n_+)/(n_- + n_+ - n_{\rm bkgd}) = 0.00\pm0.41$, where the 68\% confidence interval is obtained using the Feldman-Cousins method~\cite{PhysRevD.57.3873}.

To obtain $\mathcal{A}_{CP}$, the production, $\mathcal{A}_{\rm prod}$, reconstruction, $\mathcal{A}_{\rm reco}$, and selection, $\mathcal{A}_{\rm sel}$, asymmetries must also be accounted for.
The $D^+_s\phi$ final state is charge symmetric except for the pion from the $D_s^+$ decay.
The observed charge asymmetry in the decay modes $B^{+} \to \jpsi K^{+}$ and $B^{+} \to \Db^0 \pi^{+}$, along with the interaction asymmetry of charged kaons~\cite{LHCb-PAPER-2011-029} and the pion-detection asymmetry~\cite{Aaij:1446397} in \lhcb are used to obtain the estimate $\mathcal{A}_{\rm prod} + \mathcal{A}_{\rm reco} = (-1\pm1)\%$.  
The large $\Bsb \to D_s^+\pi^-$ sample used to determine the BDT efficiency is employed  to estimate the selection charge asymmetry yielding $\mathcal{A}_{\rm sel} = (2\pm3)\%$, where the precision is limited by the sample size.  
Finally, the \CP asymmetry is found to be
\begin{equation}
  \mathcal{A}_{CP}(\BtoDsphi) = \mathcal{A}_{\rm obs} - \mathcal{A}_{\rm prod} - \mathcal{A}_{\rm reco} - \mathcal{A}_{\rm sel} = -0.01 \pm 0.41\,({\rm stat}) \pm 0.03\,({\rm syst}), \nonumber
\end{equation} 
which is consistent with the SM expectation of no observable \CP violation.

\section{Summary}

The decay mode \BtoDsphi is seen with greater than $3\sigma$ significance.   This is the first evidence found for a hadronic annihilation-type decay of a $B^+$ meson.  
The branching fraction and \CP asymmetry for \BtoDsphi are consistent with the SM predictions.
Limits have also been set for the branching fractions of the decay modes $B^{+}_{(c)} \to D^{+}_{(s)}\Kstarz$, $B^{+}_{(c)} \to D^{+}_{(s)}\Kstarzb$ and ${B_c^{+} \to D^{+}_{s}\phi}$.  These limits are the best set to-date.

\section*{Acknowledgements}

\noindent We express our gratitude to our colleagues in the CERN accelerator
departments for the excellent performance of the LHC. We thank the
technical and administrative staff at CERN and at the LHCb institutes,
and acknowledge support from the National Agencies: CAPES, CNPq,
FAPERJ and FINEP (Brazil); CERN; NSFC (China); CNRS/IN2P3 (France);
BMBF, DFG, HGF and MPG (Germany); SFI (Ireland); INFN (Italy); FOM and
NWO (The Netherlands); SCSR (Poland); ANCS (Romania); MinES of Russia and
Rosatom (Russia); MICINN, XuntaGal and GENCAT (Spain); SNSF and SER
(Switzerland); NAS Ukraine (Ukraine); STFC (United Kingdom); NSF
(USA). We also acknowledge the support received from the ERC under FP7
and the Region Auvergne.

\bibliographystyle{LHCb}
\bibliography{main}

\end{document}